\documentclass[twocolumn]{revtex4}
\begin{document}

\title{\Large The noncovariant gauges in 3-form theories}

 \author{ Sudhaker Upadhyay\footnote {e-mail address: sudhakerupadhyay@gmail.com}}
\author{ Manoj Kumar Dwivedi\footnote {e-mail address: manojdwivedi84@gmail.com}}
\author{ Bhabani Prasad Mandal\footnote {e-mail address: bhabani.mandal@gmail.com}}

\affiliation { Department of Physics, 
Banaras Hindu University, 
Varanasi-221005, INDIA.  }

\begin{abstract}
We study the 3-form gauge theory in the context of generalized BRST formulation. We 
construct the finite field-dependent BRST (FFBRST) symmetry for such a theory. 
The generating functional for 3-form gauge theory in noncovariant gauge is obtained
from that of in covariant gauge. We further extend the results by considering 
3-form gauge theory in the context of  Batalin-Vilkovisky (BV) formulation.
\end{abstract}

\maketitle 
\section{Introduction}
Calculations in the gauge field theory have been done in variety of gauges depending on the case and convenience
of the calculation. The gauge theories in noncovariant gauges have been a subject of wide research due to certain 
advantages in these gauges \cite{iz,iz1,gri,ss,mtt,bri}. In particular, the confinement problem of QCD in 
Coulomb gauge \cite{ss}, superstring theory
in light-cone gauge \cite{mtt} are more tractable and the ultraviolet finiteness of supersymmetric Yang-Mills theory
are more transparent in noncovariant gauges \cite{bri}. 

In this present work we would like to show how noncovariant gauge formulation of Abelian 3-form gauge 
theory can be developed by making a connection of the same theory in covariant gauge.
 A 3-form field theory is subject of interest as a 3-form C field arises
naturally in M-theory. The theory of multiple M2-branes has been
used to study M5-branes and a 3-form field naturally occurs in this
theory \cite{gus,hee}. 
By considering a system of M2-branes ending on an M5-brane with a constant
3-form field turned on, the Bagger-Lambert-Gustavsson (BLG) model was used to motivate a novel
quantum geometry on the M5-brane world-volume \cite{cs}.   In particular, the action for multiple M2-branes was studied 
via BLG theory \cite{bag,bag1,gus1,mir}.
The BRST of BLG theory has been studied recently \cite{fs}.

 Another interesting relation between
multiple M2-branes and the M5-brane is the identification of the BLG action
(with Nambu-Poisson 3-bracket) as the M5-brane action with a large
world volume 3-form field \cite{pm}.
The quantization of the higher form  gauge theories
is also very important as these fields play the important roles in the excitations of
the quantized versions of strings, superstrings and other extended objects \cite{gs,pol,lt}.
Recently, the BRST and BV quantizations of Abelian 3-form gauge theory in covariant gauge have been 
studied extensively \cite{sb}.  

We show in this work how the generating functional of 3-form gauge theory in covariant 
gauge is related to that of in the noncovariant gauges by using  
  FFBRST  transformation. The FFBRST transformation of this theory is constructed by making the
infinitesimal parameter finite and 
field-dependent. The FFBRST transformations are very useful 
in connecting different effective theories and hence found many applications \cite{sdj,rb,sdj12,sud,susk,subp1,ssb,um, sd}. 
Such a generalized BRST transformation 
has also been studied in the context of BV formulation \cite{ssb,um1}.
We consider axial gauge in this work as an example of noncovariant gauge.
However, our results are general and are valid for any noncovariant 
gauge.  We further show 
the mapping between the generating functionals of  3-form gauge theories in covariant and noncovariant gauge
in BV formulation also.  

The plan of the paper is as follows. We start with a brief discussion of Abelian 3-form gauge theory in Sec. II.
The section III is devoted to generalization of BRST transformation. In Sec. IV, we show the connection between
the Abelian 3-form gauge theories in covariant and noncovariant gauges. The BV formulation for such a theory is
discussed in section V. The last section is reserved for concluding remarks. 
  
\section{Abelian 3-form gauge theories}
We start with the classical action for the Abelian 3-form gauge theory  \cite{sb} in $(1+5)$ dimensions
as 
\begin{equation}
S_0=\frac{1}{24}\int d^6x\ H_{\mu\nu\eta\chi}H^{\mu\nu\eta\chi},
\end{equation}
where the field strength (curvature) tensor in terms of totally antisymmetric tensor gauge 
field $B_{\mu\nu\eta}$ is defined as
\begin{equation}
H_{\mu\nu\eta\chi}= \partial_\mu B_{\nu\eta\chi} -\partial_\nu B_{ \eta\chi\mu} +\partial_\eta B_{\chi\mu\nu} 
-\partial_\xi B_{\mu\nu\eta}.
\end{equation}
This Lagrangian density 
is invariant under the infinitesimal gauge transformation defined as
\begin{equation}
\delta B_{\mu\nu\eta} =\partial_\mu \lambda_{\nu\eta} +\partial_\nu\lambda_{\eta\mu}+\partial_\eta\lambda_{\mu\nu},
\end{equation}
where $\lambda_{\mu\nu}$ is an arbitrary antisymmetric parameter.
To incorporate the BRST symmetry in this system, we extend the action by introducing
the following covariant gauge-fixing and ghost terms as
 \cite{sb}: 
\begin{eqnarray}
S_{gf+gh} 
&=& \int d^6x \left[ \partial_\mu B^{\mu\nu\eta}B_{\nu\eta} +
\frac{1}{2}B_{\mu\nu}\tilde B^{\mu\nu} \right.\nonumber\\
&+&\left.
(\partial_\mu \tilde c_{\nu\eta} + \partial_\nu \tilde c_{\eta\mu}
+ \partial_\eta \tilde c_{\mu\nu})\partial ^\mu 
c^{\nu\eta}\right.\nonumber\\ 
&-&  \left.(\partial_\mu\tilde \beta_\nu -\partial_\nu \tilde\beta_\mu )\partial^\mu\beta^\nu -BB_2\right.\nonumber\\
&-& \left.
   \frac{1}{2} B_1^2 +(\partial_\mu \tilde c^{\mu\nu})f_\nu -(\partial_\mu c^{\mu\nu})\tilde F_\nu \right.
\nonumber\\
&+&\left.\partial_\mu\tilde c_2 \partial^\mu c_2 
 + \tilde f_\mu f^\mu -\tilde F_\mu F^\mu \right.\nonumber\\
&+&\left. \partial_\mu\beta^\mu B_2 +\partial_\mu \phi^\mu B_1 -
\partial_\mu\tilde\beta^\mu B\right],\label{lag}
\end{eqnarray}
where antisymmetric ghost field $c_{\mu\nu}$  and antisymmetric antighost field  $\tilde c_{\mu\nu}$ are   
fermionic in nature and  the vector field 
 $\phi_\mu$, antisymmetric auxiliary fields  $B_{\mu\nu}, \tilde B_{\mu\nu}$  and  auxiliary fields  $B, 
B_1, B_2$  are bosonic in nature. Rest of the fields $f_\mu, \tilde f_\mu, F_\mu$ and $\tilde F_\mu$ are 
 auxiliary Grassmannian fields. 

The complete effective action for Abelian 3-form gauge theory is then written as
\begin{equation}
S_{eff}=S_0+S_{gf+gh}.\label{seff}
\end{equation}
This effective action ($S_{eff}$) is invariant under following BRST transformation:
\begin{eqnarray}
\delta_b B_{\mu\nu\eta} &=& -(\partial_\mu c_{\nu\eta}+\partial_\nu c_{\eta\mu} +\partial_\eta c_{\mu\nu})\delta\Lambda,\nonumber\\
\delta_b c_{\mu\nu}  &=& (\partial_\mu\beta_\nu -\partial_\nu \beta_\mu)\delta\Lambda,\
\delta_b\tilde c_{\mu\nu}=B_{\mu\nu}\delta\Lambda, \nonumber\\
\delta_b\tilde B_{\mu\nu}  &=&-(\partial_\mu f_\nu -\partial_\nu f_\mu)\delta\Lambda,   \ \ 
\delta_b\tilde\beta_\mu = -\tilde F_\mu\delta\Lambda,\nonumber\\
\delta_b\beta_\mu &=&-\partial_\mu  c_2\delta\Lambda,  \ \ \  
\delta_b F_\mu =-\partial_\mu B \delta\Lambda,\nonumber\\
\delta_b\tilde c_2 &=&B_2\delta\Lambda,\ \ \ \delta_b\tilde f_\mu =\partial_\mu B_1 \delta\Lambda,\nonumber\\
 \delta_b c_1 &=&-B\delta\Lambda,\ \ \delta_b \phi_\mu 
=-f_\mu\delta\Lambda,\nonumber\\
\delta_b \tilde c_1 &=&  B_1 \delta\Lambda,\ \ \
\delta_b {\cal M} =0,\nonumber\\
{\cal M}  &\equiv & \{c_2, f_\mu, \tilde F_\mu, B, B_1, B_2, B_{\mu\nu}\},\label{brst}
\end{eqnarray}
where $\delta\Lambda$ is infinitesimal, anticommuting and global parameter.
The gauge-fixing and ghost part of the effective action,
 $S_{gf+gh}$, is separately BRST invariant and is written in terms of BRST 
variation of the gauge-fixed fermion  ($\Psi_L$) as  
\begin{eqnarray}
S_{gf+gh}&=&s_b \Psi_L = s_b \int d^6 x[-\partial_\mu\tilde c_{\nu\eta}B^{\mu\nu\eta}-\frac{1}{2}\tilde c_2 B\nonumber\\
&+&\frac{1}{2}c_1B_2 -
\frac{1}{2}\tilde c_1 B_1 - c^{\mu\nu}
\partial_\mu\tilde \beta_\nu  - \partial_\mu \tilde c_2 \beta^\mu\nonumber\\
&+& \frac{1}{2}\tilde c_{\mu\nu} \tilde B^{\mu\nu} -F^\mu \tilde \beta_\mu 
-\tilde f^\mu\phi_\mu  ],
\end{eqnarray}
where the expression for $\Psi_L$ is given as
\begin{eqnarray}
\Psi_L &=& \int d^6 x\ \psi_L =\int d^6x [-\partial_\mu\tilde c_{\nu\eta}B^{\mu\nu\eta}-\frac{1}{2}\tilde c_2 B 
\nonumber\\
&+&\frac{1}{2}c_1B_2
 -  \frac{1}{2}\tilde c_1 B_1 - c^{\mu\nu}
\partial_\mu\tilde \beta_\nu  - \partial_\mu \tilde c_2 \beta^\mu\nonumber\\
&+& \frac{1}{2}\tilde c_{\mu\nu} \tilde B^{\mu\nu} -F^\mu \tilde \beta_\mu 
-\tilde f^\mu\phi_\mu ].\label{psi}
\end{eqnarray}
 
 The generating functional for such a theory is defined in the path integral formulation as,
 \begin{eqnarray}
 Z_{eff}=\int {\cal D}\phi\ e^{iS_{eff}},\label{zfun}
 \end{eqnarray}
 where ${\cal D}\phi$ is the path integral measure which includes all the fields $\phi$, generically.
 \section{Generalized BRST formulation of Abelian 3-form gauge theory}
The properties of the usual BRST transformation in Eq. (\ref{brst})  do not depend on whether 
the parameter $\delta\Lambda$  is (i) finite or infinitesimal, (ii) field-dependent or not, as long 
as it is anticommuting and space-time independent. These observations give us a freedom to 
generalize the BRST transformation by making the parameter, $\delta\Lambda$ finite and field-dependent without
 affecting its properties. Such generalized BRST transformation 
is known as finite field-dependent BRST (FFBRST) transformation \cite{sdj}. To generalize the BRST transformation we 
start 
by making the  infinitesimal parameter field-dependent with introduction of an arbitrary parameter $\kappa\ 
(0\leq \kappa\leq 1)$.
We allow the fields, $\phi(x,\kappa)$, to depend on  $\kappa$  in such a way that $\phi(x,\kappa =0)=\phi(x)$ and $\phi(x,\kappa 
=1)=\phi^\prime(x)$, the transformed field.

The usual infinitesimal BRST transformation, thus can be written generically as 
\begin{equation}
{d\phi(x,\kappa)}=s_{b} [\phi (x,\kappa ) ]\Theta^\prime [\phi (x,\kappa ) ]{d\kappa}
\label{diff}
\end{equation}
where the $\Theta^\prime [\phi (x,\kappa ) ]{d\kappa}$ is the infinitesimal but field-dependent parameter.
The FFBRST transformation with the finite field-dependent parameter then can be 
constructed by integrating such infinitesimal transformation from $\kappa =0$ to $\kappa= 1$, to obtain
\begin{equation}
\phi^\prime\equiv \phi (x,\kappa =1)=\phi(x,\kappa=0)+s_b[\phi(x) ]\Theta[\phi(x) ]
\label{kdep}
\end{equation}
where 
\begin{equation}
\Theta[\phi(x)]=\int_0^1 d\kappa^\prime\Theta^\prime [\phi(x,\kappa^\prime)],
\end{equation}
 is the finite field-dependent parameter \cite{sdj}. 

The FFBRST transformation is constructed in this case as follows:
\begin{eqnarray} 
&&\delta_b B_{\mu\nu\eta}= -(\partial_\mu c_{\nu\eta}+\partial_\nu c_{\eta\mu} +\partial_\eta c_{\mu\nu})
\Theta[\phi],\nonumber\\
&&\delta_b c_{\mu\nu}  = (\partial_\mu\beta_\nu -\partial_\nu \beta_\mu)\Theta[\phi],\nonumber\\
&&\delta_b\tilde c_{\mu\nu}=B_{\mu\nu}\Theta[\phi],\ \ \delta_b\tilde\beta_\mu = -\tilde F_\mu\Theta[\phi], 
\nonumber\\
&&\delta_b\tilde B_{\mu\nu}  =-(\partial_\mu f_\nu -\partial_\nu f_\mu)\Theta[\phi], 
\nonumber\\
&&\delta_b\beta_\mu =-\partial_\mu  c_2\Theta[\phi],  \ \ \  
\delta_b F_\mu =-\partial_\mu B \Theta[\phi],\nonumber\\
&&\delta_b\tilde c_2 =B_2\Theta[\phi],\ \ \ \ \delta_b\tilde f_\mu =\partial_\mu B_1 \Theta[\phi],\nonumber\\
&& \delta_b c_1 =-B\Theta[\phi],\ \ \ \delta_b \phi_\mu 
=-f_\mu\Theta[\phi],\nonumber\\
&&\delta_b \tilde c_1 =  B_1 \Theta[\phi],\ \ \ \
\delta_b \varpi =0,\nonumber\\
&&\varpi \equiv [c_2, f_\mu, \tilde F_\mu, B, B_1, B_2, B_{\mu\nu}],\label{ffbrst}
 \end{eqnarray}
Such an off-shell nilpotent BRST transformation with finite field-dependent
 parameter is the symmetry  of the effective action in Eq. (\ref{seff}). However, the 
path integral measure in Eq. (\ref{zfun}) is not invariant under such transformation as the 
BRST parameter is finite in nature.

The Jacobian of the path integral measure for such transformations is then evaluated for some 
particular choices of the finite field-dependent parameter, $\Theta[\phi(x)]$, as
\begin{eqnarray}
{\cal D}\phi^\prime &=&J(
\kappa) {\cal D}\phi(\kappa).
\end{eqnarray}
The Jacobian, $J(\kappa )$ can be replaced (within the functional integral) as
\begin{equation}
J(\kappa )\rightarrow \exp[iS_1[\phi(x,\kappa) ]]
\end{equation}
 iff the following condition is satisfied \cite{sdj}
\begin{eqnarray}
&&\int {\cal{D}}\phi (x) \;  \left [ \frac{1}{J}\frac{dJ}{d\kappa}-i\frac
{dS_1[\phi (x,\kappa )]}{d\kappa}\right ]\times \nonumber\\
&&\exp{[i(S_{eff}+S_1)]}=0 \label{mcond}
\end{eqnarray}
where $ S_1[\phi ]$ is local functional of fields.

The infinitesimal change in the $J(\kappa)$ is written as
\begin{equation}
\frac{1}{J}\frac{dJ}{d\kappa}=-\int d^6y\left [\pm \delta_b \phi (y,\kappa )\frac{
\partial\Theta^\prime [\phi (y,\kappa )]}{\partial\phi (y,\kappa )}\right],\label{jac}
\end{equation}
where $\pm$ sign refers to whether $\phi$ is a bosonic or a fermionic field.

By constructing appropriate $\Theta$, we can change $S_1$ in such a manner that $S_{eff}+S_1$ becomes a new effective 
action. 
\section{  3-form gauge theory in  noncovariant gauge}
To obtain the generating functional for 3-form gauge theory in noncovariant gauge
we construct the infinitesimal field-dependent  parameter as
\begin{eqnarray}
\Theta^\prime &=& i\gamma\int d^6y [-\tilde c_{\nu\eta}\partial_\mu B^{\mu\nu\eta} +
\tilde c_{\nu\eta}\eta_\mu B^{\mu\nu\eta}\nonumber\\
&+&c_{\mu\nu}\partial^\mu\tilde\beta^\nu - c_{\mu\nu}\eta^\mu\tilde\beta^\nu -\tilde c_2\partial_\mu \beta^\mu
\nonumber\\
&+& \tilde c_2\eta_\mu \beta^\mu ],\label{thet}
\end{eqnarray} 
where $\gamma$ is an arbitrary constant parameter.

Using Eq. (\ref{jac}), we calculate the infinitesimal change in Jacobian of functional integral as
\begin{eqnarray}
\frac{1}{J}\frac{dJ}{d\kappa}&=&-i\gamma\int d^6y\left [B_{\nu\eta}\partial_\mu B^{\mu\nu\eta} -
B_{\nu\eta}\eta_\mu B^{\mu\nu\eta}\right.\nonumber\\
&+&\left. \partial^\mu(\partial_\mu c_{\nu\eta} +\partial_\nu c_{\eta\mu} +\partial_\eta c_{\mu\nu})\tilde c^{\nu\eta}
\right.\nonumber\\
&-&\left. \eta^\mu(\partial_\mu c_{\nu\eta} +\partial_\nu c_{\eta\mu} +\partial_\eta c_{\mu\nu})\tilde c^{\nu\eta}
\right.\nonumber\\
&-&\left.(\partial_\mu\tilde \beta_\nu -\partial_\nu\tilde \beta_\mu )\partial^\mu\beta^\nu +c_{\mu\nu}
\partial^\mu\tilde F^\nu \right.\nonumber\\
&+&\left.(\eta_\mu\tilde \beta_\nu -\eta_\nu\tilde \beta_\mu )\partial^\mu\beta^\nu -c_{\mu\nu}
\eta^\mu\tilde F^\nu \right.\nonumber\\
&+&\left. B_2\partial_\mu\beta^\mu -B_2\eta_\mu\beta^\mu -\tilde c_2\partial_\mu\partial^\mu c_2\right.\nonumber\\
&+&\left. \tilde c_2\eta_\mu\partial^\mu c_2\right].
\end{eqnarray}
The Jacobian $J$ can be written as $e^{iS_1}$ when the condition (\ref{mcond}) is satisfied.
We make the following ansatz for functional $S_1$ in this case:
\begin{eqnarray}
S_1 &=&\int d^6 x [\xi_1 (\kappa ) B_{\nu\eta}\partial_\mu B^{\mu\nu\eta } +
\xi_2 (\kappa ) B_{\nu\eta } \eta_\mu B^{\mu\nu\eta } \nonumber\\
&+& \xi_3 (\kappa )\partial^\mu(\partial_\mu c_{\nu\eta} +\partial_\nu c_{\eta\mu} +\partial_\eta c_{\mu\nu} )
\tilde c^{\nu\eta}\nonumber\\
&+& \xi_4 (\kappa )\eta^\mu(\partial_\mu c_{\nu\eta} +\partial_\nu c_{\eta\mu} +\partial_\eta c_{\mu\nu} )
\tilde c^{\nu\eta}\nonumber\\
&+&\xi_5 (\kappa) (\partial_\mu\tilde \beta_\nu -\partial_\nu\tilde \beta_\mu )\partial^\mu\beta^\nu \nonumber\\
&+&\xi_6 (\kappa) (\eta_\mu\tilde \beta_\nu -\eta_\nu\tilde \beta_\mu )\partial^\mu\beta^\nu \nonumber\\
&+& \xi_7 (\kappa ) c_{\mu\nu} \partial^\mu \tilde F^{\nu} +\xi_8 (\kappa ) c_{\mu\nu} \eta^\mu \tilde F^{\nu}
\nonumber\\
&+& \xi_9 (\kappa) B_2\partial_\mu \beta^\mu +\xi_{10}(\kappa) B_2\eta_\mu \beta^\mu \nonumber\\
&+&
\xi_{11}(\kappa) \tilde c_2\partial_\mu\partial^\mu c_2 +\xi_{12}(\kappa) \tilde c_2\eta_\mu\partial^\mu c_2 
].\label{s1}
\end{eqnarray}
Now, the condition (\ref{mcond}) yields
\begin{eqnarray}
&&\int {\cal{D}}\phi (x) \left 
[(\xi_1^\prime +\gamma) B_{\nu\eta}\partial_\mu B^{\mu\nu\eta }\right.\nonumber\\
& +&\left. (\xi_2^\prime -\gamma) B_{\nu\eta } \eta_\mu B^{\mu\nu\eta }\right. \nonumber\\
&+& \left.(\xi_3^\prime +\gamma)\partial^\mu(\partial_\mu c_{\nu\eta} +\partial_\nu c_{\eta\mu}
 +\partial_\eta c_{\mu\nu} ) \tilde c^{\nu\eta}\right.\nonumber\\
&+&\left. (\xi_4^\prime -\gamma)
\eta^\mu(\partial_\mu c_{\nu\eta} +\partial_\nu c_{\eta\mu} +\partial_\eta c_{\mu\nu} )
\tilde c^{\nu\eta}\right.\nonumber\\
&+&\left. (\xi_5^\prime -\gamma)(\partial_\mu\tilde \beta_\nu -\partial_\nu\tilde \beta_\mu )
\partial^\mu\beta^\nu\right. \nonumber\\
&+&\left.(\xi_6^\prime +\gamma)(\eta_\mu\tilde \beta_\nu -\eta_\nu\tilde \beta_\mu )\partial^\mu\beta^\nu 
\right. \nonumber\\
&+&\left.(\xi_7^\prime +\gamma) c_{\mu\nu} \partial^\mu \tilde F^{\nu} 
+ (\xi_8^\prime -\gamma) c_{\mu\nu} \eta^\mu \tilde F^{\nu}\right.\nonumber\\
&+&\left.( \xi_9^\prime +\gamma) B_2\partial_\mu \beta^\mu +(\xi_{10}^\prime -\gamma) B_2\eta_\mu \beta^\mu
 \right.\nonumber\\
&+& \left.(
\xi_{11}^\prime -\gamma) \tilde c_2\partial_\mu\partial^\mu c_2 +(\xi_{12}^\prime +\gamma) \tilde c_2
\eta_\mu\partial^\mu c_2 \right.
\nonumber\\
&-&\left. \beta_{\nu\eta}\partial_\mu (\partial^\mu c^{\nu\eta}+\partial^\nu c^{\eta\mu}+\partial^\eta c^{\mu\nu})
\Theta'(\xi_1-\xi_3)\right.\nonumber\\
 &-&\left. \beta_{\nu\eta}\eta_\mu (\partial^\mu c^{\nu\eta}+\partial^\nu c^{\eta\mu}+\partial^\eta c^{\mu\nu})
\Theta'(\xi_2 -\xi_4)\right.\nonumber\\
&-&\left. (\partial_\mu\tilde F_\nu -\partial_\nu\tilde F_\mu)\partial^\mu\beta^\nu \Theta' (\xi_5 +\xi_7 )
\right.\nonumber\\
&-&\left. (\eta_\mu\tilde F_\nu -\eta_\nu\tilde F_\mu)\partial^\mu\beta^\nu \Theta' (\xi_6 +\xi_8 )\right.\nonumber\\
&-&\left. B_2\partial_\mu\partial^\mu c_2\Theta' (\xi_9 +\xi_{11})\right.\nonumber\\
&-&\left. B_2\eta_\mu\partial^\mu c_2\Theta' (\xi_{10} +\xi_{12})  \right] =0.\label{cond}
\end{eqnarray}
The nonlocal ($\Theta'$ dependent) terms will  vanish if the following conditions are satisfied:
\begin{eqnarray}
&&\xi_1-\xi_3 =\xi_2 -\xi_4=\xi_5 +\xi_7=\nonumber\\
 &&\xi_6 +\xi_8=\xi_9 +\xi_{11}=\xi_{10} +\xi_{12} =0.
\end{eqnarray}
Equating the L.H.S. and R.H.S. of the Eq. (\ref{cond}) we get the following differential equations:
\begin{eqnarray}
\xi_1^\prime +\gamma &=&0,\ \ \xi_2^\prime -\gamma =0, \nonumber\\
\xi_3^\prime +\gamma &=&0,\ \ \xi_4^\prime -\gamma  =0, \nonumber\\
\xi_5^\prime -\gamma &=&0,\ \ \xi_6^\prime +\gamma  =0, \nonumber\\
\xi_7^\prime +\gamma &=&0,\ \ \xi_8^\prime -\gamma  =0, \nonumber\\
\xi_9^\prime +\gamma &=&0,\ \ \xi_{10}^\prime -\gamma  =0, \nonumber\\
\xi_{11}^\prime -\gamma &=&0,\ \ \xi_{12}^\prime +\gamma  =0.  
\end{eqnarray}
The solutions of the above differential equations  satisfying the initial conditions, i.e. 
$\xi_i(\kappa =0) =0 \ ( i=1,2, ..., 12)$,
are 
 \begin{eqnarray}
 \xi_1 =-\gamma\kappa,\ \ \xi_2 =\gamma\kappa,\ \ \xi_3 =-\gamma\kappa,\nonumber\\
 \xi_4 =\gamma\kappa,\ \ \xi_5 =\gamma\kappa,\ \ \xi_6 =-\gamma\kappa,\nonumber\\
 \xi_7 =-\gamma\kappa,\ \ \xi_8 =\gamma\kappa,\ \ \xi_9 =-\gamma\kappa,\nonumber\\
 \xi_{10} =\gamma\kappa,\ \ \xi_{11} = \gamma\kappa,\ \ \xi_{12} =-\gamma\kappa.
 \end{eqnarray}
 Plugging back these values, the expression of $S_1$ given in Eq. (\ref{s1}) becomes 
\begin{eqnarray}
S_1 &=&\int d^6 x [-\gamma\kappa B_{\nu\eta}\partial_\mu B^{\mu\nu\eta } +
\gamma\kappa B_{\nu\eta } \eta_\mu B^{\mu\nu\eta } \nonumber\\
&-&  \gamma\kappa\partial^\mu(\partial_\mu c_{\nu\eta} +\partial_\nu c_{\eta\mu} +\partial_\eta c_{\mu\nu} )
\tilde c^{\nu\eta}\nonumber\\
&+& \gamma\kappa\eta^\mu(\partial_\mu c_{\nu\eta} +\partial_\nu c_{\eta\mu} +\partial_\eta c_{\mu\nu} )
\tilde c^{\nu\eta}\nonumber\\
&+&\gamma\kappa (\partial_\mu\tilde \beta_\nu -\partial_\nu\tilde \beta_\mu )\partial^\mu\beta^\nu \nonumber\\
&-&\gamma\kappa (\eta_\mu\tilde \beta_\nu -\eta_\nu\tilde \beta_\mu )\partial^\mu\beta^\nu \nonumber\\
&-& \gamma\kappa c_{\mu\nu} \partial^\mu \tilde F^{\nu} +\gamma\kappa c_{\mu\nu} \eta^\mu \tilde F^{\nu}
\nonumber\\
&-& \gamma\kappa B_2\partial_\mu \beta^\mu +\gamma\kappa B_2\eta_\mu \beta^\mu \nonumber\\
&+&
\gamma\kappa \tilde c_2\partial_\mu\partial^\mu c_2 -\gamma\kappa \tilde c_2\eta_\mu\partial^\mu c_2 
].\label{s2}
\end{eqnarray} 
We consider  the arbitrary constant $\gamma =1$ in the above expression without any loss of generality and by
 adding the expression (\ref{s2}) at $\kappa =1$ to the effective action for 3-form gauge theory in covariant gauge,
 we get
 \begin{eqnarray}
 S_{eff} &+&S_1 (\kappa =1) =\int d^6 x\left [\frac{1}{24}H_{\mu\nu\eta\chi}H^{\mu\nu\eta\chi}\right.\nonumber\\
 &+&\left.\eta_\mu B^{\mu\nu\eta}B_{\nu\eta} +
\frac{1}{2}B_{\mu\nu}\tilde B^{\mu\nu} \right.\nonumber\\
&+&\left.
(\partial_\mu \tilde c_{\nu\eta} + \partial_\nu \tilde c_{\eta\mu}
+ \partial_\eta \tilde c_{\mu\nu})\eta^\mu 
c^{\nu\eta}\right.\nonumber\\ 
&-&  \left.(\eta_\mu\tilde \beta_\nu -\eta_\nu \tilde\beta_\mu )\partial^\mu\beta^\nu -BB_2\right.\nonumber\\
&-& \left.
   \frac{1}{2} B_1^2 +(\partial_\mu \tilde c^{\mu\nu})f_\nu +  c_{\mu\nu}\eta^\mu\tilde F^\nu \right.
\nonumber\\
&-&\left.\tilde c_2 \eta_\mu\partial^\mu c_2 
 + \tilde f_\mu f^\mu -\tilde F_\mu F^\mu \right.\nonumber\\
&+&\left. \eta_\mu\beta^\mu B_2 +\partial_\mu \phi^\mu B_1 -
\partial_\mu\tilde\beta^\mu B\right].
 \end{eqnarray}
This is nothing but the 3-form effective action in noncovariant gauge.

We end this this section by making the  comment that  the noncovariant 
gauge formulation of 3-form gauge theory is obtained from covariant gauge formulation of the same
through a FFBRST transformation with appropriate finite field dependent 
parameter. Even though we have established the connection between generating functionals 
in Lorentz (covariant) gauge and in axial (noncovariant) gauge our formulation is valid
for the connection between generating functionals in any covariant and in any noncovariant gauges.

\section{Mapping of covariant and noncovariant gauges in 3-form gauge theory: BV formulation}
In this section, we consider the BV formulation for Abelian 3-form gauge theory to reestablish the results 
of the previous section. For this purpose we express the 
generating functional in Eq. (\ref{zfun}) in field/antifield formulation by introducing the
antifield $\phi^\star $ corresponding to each generic field $\phi$ with opposite statistics, as  
\begin{eqnarray}
Z_{eff} &=&\int {\cal D}\phi \exp\left[i\int d^6x\left\{\frac{1}{24}F_{\mu\nu\eta\chi}F^{\mu\nu\eta\chi}
\right.\right. \nonumber\\
 & -& \left.\left. 
B_{\mu\nu\eta}^\star (
\partial^\mu c^{\nu \eta} + \partial^\nu c^{  \eta\mu} +
\partial^\eta c^{\mu\nu }  )\right.\right. \nonumber\\
 & +& \left.\left. 
    {c}_{\mu\nu}^\star ( \partial^\mu\beta^\nu  -\partial^\nu\beta^\mu )+ \tilde{c}_{\mu\nu}^\star 
 B^{\mu\nu} \right.\right. \nonumber\\
 & -& \left.\left.\tilde B_{\mu\nu}^\star (\partial^\mu f^\nu   -\partial^\nu f^\mu  ) 
-\beta_\mu^\star \partial^\mu c_2  -\tilde \beta_\mu^\star \tilde F ^\mu \right.\right. \nonumber\\
 & -& \left.\left.
 F_\mu^\star  \partial^\mu B  + \tilde  f_\mu^\star  \partial ^\mu B_1 +
 \tilde  c_2^\star   B_2  
 +\tilde c_1^\star  B_1 \right.\right. \nonumber\\
 & -& \left.\left. c_1^\star  B 
-\phi_\mu ^\star   f^\mu \right\}\right ].
\end{eqnarray}
This further can be written in compact form as
\begin{equation}
Z_{eff} = \int {\cal D}\phi\ e^{iW_{\Psi_L}(\phi,\phi^\star)},
\end{equation}
where $W_{\Psi_L}(\phi,\phi^\star)$ is an extended action for Abelian 3-form gauge theory in the covariant 
gauge corresponding to the gauge-fixed fermion $\Psi_L $ (Eq. (\ref{psi})) having Grassmann  parity 1 and ghost 
number ${-1}$.

The generating functional $Z_{eff}$ does not depend on the choice of gauge-fixed fermion \cite{ht}.
This extended quantum action, $W_{\Psi_L}(\phi,\phi^\star)$, satisfies  certain rich mathematical
relation which is called  quantum master equation \cite{wei} and is given by
\begin{equation}
\Delta e^{iW_{\Psi_L}[\phi, \phi^\star ]} =0,\ \
 \Delta\equiv \frac{\partial_r}{
\partial\phi}\frac{\partial_r}{\partial\phi^\star } (-1)^{\epsilon
+1}.
\label{mq}
\end{equation}
The antifields $\phi^\star(=\frac{d\psi_L}{d\phi})$ corresponding to the each generic 
field $\phi$ for this particular theory are obtained from the gauge-fixed fermion as 
\begin{eqnarray}
&&B_{\mu\nu\eta}^\star=-\partial_\mu \tilde c_{\nu\eta},\ \
  c_{\mu\nu}^\star  = -\partial_\mu\tilde\beta_\nu,\ \
 \tilde B_{\mu\nu}^\star = \frac{1}{2}\tilde c_{\mu\nu},\nonumber\\
&&\tilde c_{\mu\nu}^\star =\frac{1}{2}\tilde B_{\mu\nu} +
\partial^\eta B_{\mu\nu\eta},\ \
 \beta_{\mu}^\star =- \partial_\mu\tilde c_2,\nonumber\\
&&\tilde\beta_{\mu}^\star =-F_\mu +\partial^\nu c_{\nu\mu},\ \  
 F_{\mu\star} =-\tilde\beta_\mu,\ \
\tilde f_{\mu}^\star  =-\phi_\mu,\nonumber\\ 
&&\tilde c_2^{\star}=-\frac{1}{2}B+\partial_\mu\beta^\mu,\ \
  c_1^{\star} =\frac{1}{2}B_2,\ \
 \tilde c_1^{\star} =-\frac{1}{2}B_1,\nonumber\\
&&\phi_{\mu}^\star =-\tilde f_\mu,\ \
 B^{\star} =-\frac{1}{2}\tilde c_2,\ \
  B_1^{\star} =-\frac{1}{2}\tilde c_1,\nonumber\\
&&B_2^{\star} = \frac{1}{2}c_1,\ \
 \{  B_{\mu\nu}^\star, \tilde F_{\mu}^\star, f_{\mu }^\star,   c_2^{\star}\}=0.
\end{eqnarray}

Now, we construct the field-dependent parameter written in field/antifield
formulation as
\begin{eqnarray}
\Theta^\prime &=& i\gamma\int d^6y [-B^\star_{\mu\nu\eta} B^{\mu\nu\eta} +\bar B^\star_{\mu\nu\eta} B^{\mu\nu\eta} 
\nonumber\\
&-&c^\star_{\mu\nu}c^{\mu\nu} +\bar c^\star_{\mu\nu}c^{\mu\nu} -\beta_\mu^\star \beta^\mu
+\bar\beta_\mu^\star \beta^\mu ],\label{th}
\end{eqnarray} 
where bar fields are the antifields corresponding to the fields satisfying axial (noncovariant)
 gauge-fixing condition.  
Under the FFBRST transformation with such field dependent parameter, the path integral measure is not invariant 
and give rise to a factor which is written as 
$e^{iS_1}$, and the expression for functional $S_1$ is calculated using Eqs. (\ref{mcond}), (\ref{jac}) and  
(\ref{th}) as
 \begin{eqnarray}
S_1 &=&\int d^6 x [  \kappa B_{\mu\nu\eta}^\star (
\partial^\mu c^{\nu \eta} + \partial^\nu c^{  \eta\mu} +
\partial^\eta c^{\mu\nu }  )\nonumber\\
 &-& \kappa \bar B_{\mu\nu\eta}^\star (
\partial^\mu c^{\nu \eta} + \partial^\nu c^{  \eta\mu} +
\partial^\eta c^{\mu\nu }  )\nonumber\\
& -& \kappa  
    {c}_{\mu\nu}^\star ( \partial^\mu\beta^\nu  -\partial^\nu\beta^\mu )+ \kappa  
    \bar{c}_{\mu\nu}^\star ( \partial^\mu\beta^\nu \nonumber\\
 &-& \partial^\nu\beta^\mu )
- \kappa  \tilde{c}_{\mu\nu}^\star  B^{\mu\nu}+ \kappa \bar{ \tilde{c}}_{\mu\nu}^\star 
 B^{\mu\nu} \nonumber\\
& +& \kappa  \tilde B_{\mu\nu}^\star (\partial^\mu f^\nu    
  - \partial^\nu f^\mu  )
  -  \kappa \bar {\tilde B}_{\mu\nu}^\star (\partial^\mu f^\nu  \nonumber\\
&-&\partial^\nu f^\mu  ) 
+ \kappa\beta_\mu^\star \partial^\mu c_2 
- \kappa\bar\beta_\mu^\star \partial^\mu c_2  + \kappa\tilde \beta_\mu^\star \tilde F ^\mu  \nonumber\\
&-& \kappa\bar{\tilde \beta}_\mu^\star \tilde F ^\mu
+ \kappa F_\mu^\star  \partial^\mu B  
- \kappa \bar F_\mu^\star  \partial^\mu B \nonumber\\ &-& \kappa \tilde  f_\mu^\star  \partial ^\mu B_1
+ \kappa \bar{\tilde  f}_\mu^\star  \partial ^\mu B_1 
- \kappa
 \tilde  c_2^\star   B_2  \nonumber\\
 &+& \kappa
 \bar{\tilde  c}_2^\star   B_2 
 - \kappa\tilde c_1^\star  B_1 + \kappa\bar{\tilde c}_1^\star  B_1+ \kappa c_1^\star  B \nonumber\\
 & -& \kappa \bar c_1^\star  B 
+ \kappa\phi_\mu ^\star   f^\mu - \kappa \bar\phi_\mu ^\star   f^\mu ]. 
\end{eqnarray}
The transformed generating functional in BV formulation is given by
\begin{eqnarray}
Z_{eff}^\prime &=&\int {\cal D}\phi\ e^{i\{ W_{\Psi_L}+S_1(\kappa =1)\}}, \nonumber\\
&\equiv &\int {\cal D}\phi\ e^{i W_{\Psi_A}}.
\end{eqnarray}
$Z_{eff}^\prime$ is the generating functional for Abelian 3-form gauge theory in the axial gauge
with gauge-fixing fermion
\begin{eqnarray}
\Psi_A &=&\int d^6x \left[ -\eta_\mu\tilde c_{\nu\eta}B^{\mu\nu\eta}-\frac{1}{2}\tilde c_2 B
+\frac{1}{2}c_1B_2\right.\nonumber\\ 
&-&\left.
\frac{1}{2}\tilde c_1 B_1 - c^{\mu\nu}
\eta_\mu\tilde \beta_\nu  - \eta_\mu \tilde c_2 \beta^\mu +\frac{1}{2}\tilde c_{\mu\nu} \tilde B^{\mu\nu} 
\right.\nonumber\\
&-& \left. F^\mu \tilde \beta_\mu 
-\tilde f^\mu\phi_\mu \right].
\end{eqnarray}
The extended action $ W_{\Psi_A}$ for 3-form gauge theory in axial gauge also satisfies  
the quantum master equation (\ref{mq}). This result is also true for the connection of generating functionals 
in any covariant and noncovariant gauges in BV formulation.

\section{Conclusions}
The 3-form gauge theory in noncovariant gauge has been developed with the help of the finite 
field dependent BRST transformation. Usual BRST transformations have 
been generalized to the case of Abelian 3-form gauge theory in covariant gauge.
The generating functional for this theory is shown to be connected to that of in the noncovariant (axial)
gauge through FFBRST transformation. We established this connection by constructing
appropriate finite field dependent BRST parameter.
However, the various noncovariant gauges like the Coulomb gauge, the light-cone gauge, the planer gauge
and the temporal gauge can also be studied under such formulation. 
Thus, our formulation enables to study the 3-form gauge theory in noncovariant 
gauges by this connection. We further extend these results in the context 
of BV formulation. It will be interesting to see how the different Green functions are
related in different gauges for this theory. 
Further analysis of FFBRST transformation
of quantum gravity, BLG theory, ABJM theory and higher derivative field
theory will be subject of interest.

\begin{acknowledgments}
SU gratefully acknowledges the financial support from the Council of Scientific and Industrial Research
(CSIR), India, under the SRF scheme. 
\end{acknowledgments}


\begin{thebibliography}{99}
\bibitem{iz} George Leibbrandt, {\it Rev. Mod. Phys.} {\bf 59}, 1067 (1987).
\bibitem{iz1} George Leibbrandt,   {\it Noncovariant Gauges:
Quantization of Yang-Mills and Chern-Simons Theory in Axial-Type Gauges}, (World Scientific, 1994).
\bibitem{gri} V. N. Gribov,  Nucl. Phys. {\bf B 139}, 1 (1978).
\bibitem{ss} A. P. Szczepaniak and E. S. Swanson,  {Phys. Rev.}  {\bf D 65}, 025012 (2001).
\bibitem{mtt} R. R. Metsaev, C. B. Thorn and  A. A. Tseytlin, Nucl. Phys. {\bf B 596}, 151 (2001).
\bibitem{bri} L. Brink, O. Lindgren and B. E. W. Nilsson, Phys. Lett. {\bf B 123}, 323 (1983).
\bibitem{gus} A. Gustavsson, JHEP {\bf 1201}, 057 (2012).
\bibitem{hee} H.-C. Kim, S.
Kim, E. Koh, K. Lee, S. Lee, JHEP {\bf 12 }, 031 (2011).
\bibitem{cs} C.-S. Chu and D. J.
Smith, JHEP. {\bf 0904}, 097 (2009).
\bibitem{bag} J. Bagger and N. Lambert, JHEP. {\bf 0802}, 105 (2008).
\bibitem{bag1} J. Bagger and N. Lambert, Phys. Rev. {\bf D 77}, 065008 (2008).
\bibitem{gus1} A. Gustavsson, JHEP {\bf 0804}, 083 (2008).
\bibitem{mir} M. Faizal, JHEP {\bf 1204}, 017 (2012). 
\bibitem{fs} M. Faizal, B. P. mandal and S. Upadhyay, arXiv:1212.5653 [hep-th].
 \bibitem{pm} P.-M. Ho, Chin. J. Phys. {\bf 48}, 1 (2010)
\bibitem{gs} M. B. Green, J. H. Schwarz and E. Witten, {\it Superstring Theory} Vols 1 and 2
(Cambridge University Press: Cambridge, 1987).
\bibitem{pol} J. Polchinski, {\it String Theory} Vols 1 and 2
(Cambridge University Press: Cambridge, 1998).
\bibitem{lt} D. Lust, S. Theisen, {\it Lectures in String Theory} (Springer-Verlag: New York, 1989).


\bibitem{sb} S. Upadhyay and B. P. Mandal, Eur. Phys. J. {\bf C 72},   2059 (2012).
\bibitem{sdj} S. D. Joglekar and B. P. Mandal,   {Phys. Rev.}  {\bf{D 51}}, 1919 (1995).
\bibitem{rb}  R. Banerjee and B. P. Mandal,  Phys. Lett.  {\bf B  27}, {488} (2000).
 \bibitem{sdj12}S. D. Joglekar and B. P. Mandal, Int.J. Mod.Phys. {\bf A 17},  1279 (2002).
\bibitem{sud} S. Upadhyay and B. P. Mandal, Eur. Phys. Lett. {\bf 93}, 31001 (2011).
\bibitem{susk}   S. Upadhyay,   S. K. Rai and B. P. Mandal,  J. Math. Phys. \textbf{52}, (2011) {022301}.
 \bibitem{subp1}    S. Upadhyay and B. P. Mandal,  Eur. Phys. Lett. \textbf{ 93}, 
{31001} (2011);   Mod. Phys. Lett. \textbf{A 40}, { 3347} (2010) .
\bibitem{ssb}  B. P. Mandal,  S. K. Rai and S.  Upadhyay,
Eur. Phys. Lett. \textbf{ 92}, {21001} (2010).
\bibitem{um}  S. Upadhyay and B. P. Mandal,   Eur. Phys. J. \textbf{C 72},  2065 (2012).
\bibitem{sd}  S. Upadhyay and B. P. Mandal, Annals of Physics {\bf 327} 2885(2012).
\bibitem{um1}  S. Upadhyay and B. P. Mandal,   AIP Conf. Proc. {\bf 1444}, 213 (2012).
 \bibitem{ht} M. Henneaux and C. Teitelboim, {\it{ Quantization of gauge
systems}} (Princeton, USA: Univ. Press, 1992).
\bibitem{wei} S. Weinberg, {\it{ The quantum theory of fields, Vol-II: Modern
applications}} (Cambridge, UK Univ. Press, 1996).
\end{thebibliography}
\end{document}